\documentclass[twocolumn,showpacs,prb,graphics,graphicx]{revtex4-1}
\usepackage{bm}
\usepackage[pdftex]{graphicx}
\usepackage{dcolumn}

\begin{document}

\title{
Interaction of Josephson junction and distant vortex in narrow thin-film superconducting strips}
\author{V. G. Kogan}
\affiliation{Ames Laboratory - DOE and Department of Physics, Iowa State University,
Ames, IA 50011, USA}
\author{R. G. Mints}
\affiliation{School of Physics and Astronomy, Raymond
and Beverly Sackler Faculty of Exact Sciences, 
Tel Aviv University, Tel Aviv 69978, Israel} 
\begin{abstract}
 The phase difference between the banks of an edge-type planar Josephson junction  crossing the narrow thin-film strip depends on wether or not  vortices are present
 in the junction banks. For a vortex close to the junction this effect has been seen by  Golod,  Rydh, and  Krasnov,  \prl {\bf 104},   227003 (2010), who showed that the vortex may turn the junction into $\pi$-type. It is shown here that even if the vortex  is far away from the junction, it still changes the 0-junction to $\pi$-junction when situated close to the strip edges. Within the approximation used, the latter effect is independent of the vortex-junction separation, a    manifestation of  topology of the vortex phase which extends to macroscopic distances of superconducting coherence.
 \end{abstract}

\date{\today}

\maketitle

It is long known that Abrikosov vortices in the vicinity of Josephson junctions affect the junction properties.\cite{Ustinov,Fistul} Recent experiments with a vortex trapped in one of the banks of an edge-type  planar junction in a thin-film superconducting strip  showed that the vortex causes an extra phase difference on the junction that depends on the vortex position.\cite{Krasnov} This effect is strong in particular when the vortex is close to the junction, the situation when  the junction character is changed from the conventional ``zero"-type behavior to that of the $\pi$-junction. In this communication we show that this can happen even if the vortex  is far away from the junction but close to the strip edges. This   is  manifestation of the phase coherence on macroscopic distances in superconductors and of the topology of the vortex phase. The effect can be utilized  for manipulating Josephson currents by  controlling  the far-away vortex position. \\

\begin{figure}[h]
\begin{center}
 \includegraphics[width=7cm] {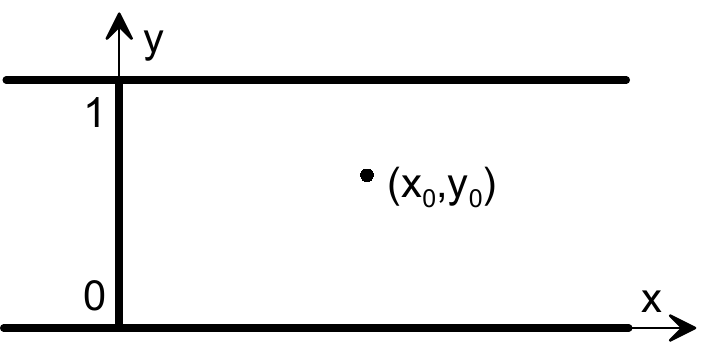}
\caption{  The superconducting thin-film strip with a Josephson junction at $x=0$   and a vortex at $(x_0,y_0)$. The width $W$ is used as a unit length.   
}
\label{fig1}
\end{center}
\end{figure}

1. $\bm{The\,\,\, problem.}$ Consider a  thin-film strip of a width $W$ with an edge-type  Josephson junction across the strip. The strip is narrow: $W \ll \Lambda=2\lambda^2/d$  where $\lambda$ is the London penetration depth of the film material and $d$ is the film thickness. Choose $x$ along the strip and $y$ across so that $0<y<W$ and the junction is at $x=0$. Let a vortex be pinned at $\bm r_0=(x_0,y_0)$. The problem is to evaluate the phase of the vortex along the bank $x=+0$ of the junction, Fig.\ref{fig1}.   

The London equation  integrated over the film thickness is
\begin{equation}\label{eqn_04}
\frac{c} {2\pi\Lambda }h_z + {\rm curl}_z\,{\bm g}
=\frac{c \phi_0} {2\pi\Lambda } \,\delta(\bm r-\bm r_0) \,,
\label{London}
\end{equation}
where  $h_z$ is the field component normal to the film and $\bm g(x,y)$ is the sheet current density. 
  Since div\,$\bm g=0$, one can look for ${\bm g} ={\rm curl}\,S{\hat{\bm z}}$, where
$S(x,y)$ is the scalar stream function. For large $\Lambda$, the first term in (\ref{London}) is small and we have
\begin{equation}\label{eqn_04}
\nabla^2S
=-\frac{c \phi_0} {2\pi\Lambda } \,\delta(\bm r-\bm r_0) \,.
\label{Poisson}
\end{equation}
This is, in fact, a Poisson equation for a linear ``charge" $c \phi_0/8\pi^2\Lambda$ at $\bm r_0$ so that the problem is equivalent  to that in the two-dimensional (2D) electrostatics.
 
 Moreover,  the current is expressed either in terms of the gauge invariant phase $\varphi$ or via the stream function $S$:  
${\bm g}=-(c\phi_0/4\pi^2\Lambda)\nabla \varphi 
={\rm curl}\,S {\bm z}$. This  relation written in components shows that 
$(4\pi^2\Lambda/c\phi_0)S({\bm r})$ and $\varphi({\bm r})$
are the real and imaginary parts of an analytic function.  
 
\begin{figure}[h]
\begin{center}
 \includegraphics[width=6cm] {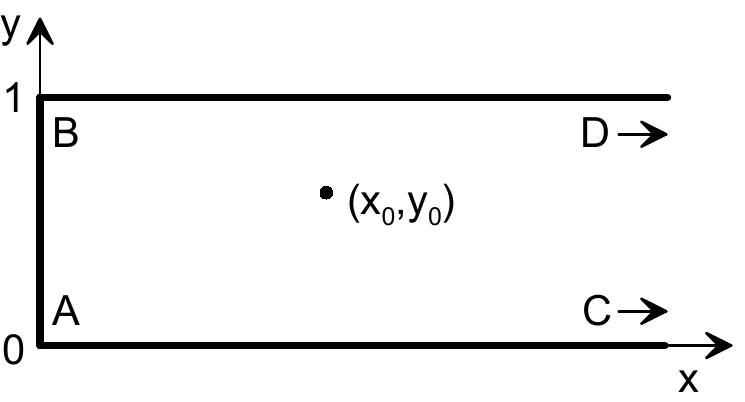}
   \includegraphics[width=6cm] {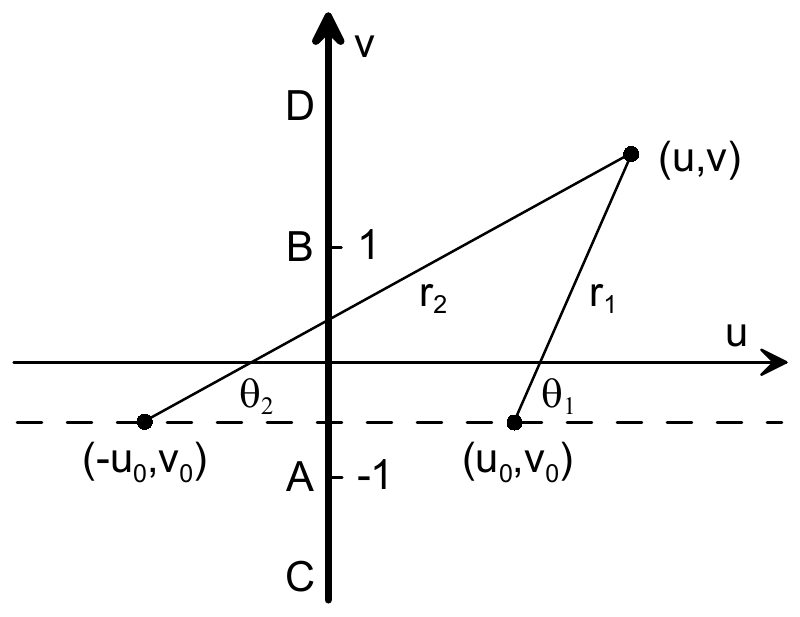}
\caption{  Upper panel: the half-strip of the width 1 with a vortex at $(x_0,y_0)$. Eqs.\,(\ref{eq3}) or (\ref{eq4}) map the half-strip in the $(x,y)$ plane onto the half-plane $u>0$ (lower panel) with the vortex at $(u_0,v_0)$; the points A,B,C,D on both planes are shown. 
}
\label{fig2}
\end{center}
\end{figure}

The sheet current normal to the strip
edges $y=0,W$ is zero. Besides,   one can disregard Josephson tunneling currents relative to those of the vortex, i.e. to set $g_x(0,y)=0$ as well. These boundary conditions imply that $S$ is a constant along the edges of the half-strip; one can choose this constant as zero.  Thus, the problem is formally equivalent to the 2D problem of electrostatic potential $S$ due to a  linear charge at $\bm r_0$ on a half-strip with grounded edges, see the upper panel of Fig.\,\ref{fig2}. \\

2. $\bm{Conformal\,\,\, mapping.}$ This problem can be solved by  conformal mapping of the half-strip onto a half-plane for which the electrostatic field is easily found.\cite{remark0} The relation    
 \begin{eqnarray}
u+i v=-i\cosh \pi(x+i y) 
 \label{eq3}
\end{eqnarray}
 transforms the   half-plane $u>0$ to the half-strip of our interest as shown in Fig.\,\ref{fig2}.  In real terms this transformation  reads:
 \begin{eqnarray}
u =\sinh \pi x\,\sin \pi y  \,, \qquad 
v=-\cosh \pi x\,\cos \pi y\,.  
 \label{eq4}
\end{eqnarray}
 In particular,   the vortex position  $(x_0,y_0)$  transforms to the point  $(u_0,v_0)$:
  \begin{eqnarray}
u_0 =\sinh \pi x_0\,\sin \pi y_0 \,, \quad 
v_0=-\cosh \pi x_0\,\cos \pi y_0 \,. \qquad 
 \label{eq5}
\end{eqnarray}
 
  The complex potential $F(u,v)$ for a linear charge (vortex) at $w_v=u_0+iv_0$ at the half plane $ u>0 $ near the grounded plane at $u=0$ is
 \begin{eqnarray}
F=2 q\ln\frac{w-w_v}{w-w_{av}} = 2q\left[\ln\frac{r_1}{r_2}+i (\theta_1-\theta_2)\right] 
 \label{eq6}
\end{eqnarray}
where $w=u+iv$, $w_{av}=-u_0+i\,v_0$ is the position of fictitious antivortex. 
$r_{1,2}$ and $\theta_{1,2}$ are the corresponding moduli and phases, see the lower panel of Fig.\,\ref{fig2}; $q$ is the linear ``charge" $c\phi_0/8\pi^2\Lambda$.   Clearly,
  \begin{eqnarray}
r_1 &=&\sqrt{(u-u_0)^2+(v-v_0)^2}  \,, \nonumber\\
r_2&=&\sqrt{(u+u_0)^2+(v -v_0)^2}  \,,  
 \label{eq7}
\end{eqnarray}
and 
  \begin{eqnarray}
\theta_1 - \theta_2=\tan^{-1}\frac{v-v_0}{u-u_0} -  \tan^{-1}\frac{v-v_0}{u+u_0}
  \,.  
 \label{eq8}
\end{eqnarray}
The phase in the plane $(x,y)$ is obtained by substitution   of Eqs.\,(\ref{eq4}) and (\ref{eq5})	 in (\ref{eq8}).	 		
 
  We are interested in the vortex phase at the junction bank $x=+0, \,\,0<y<1$, which corresponds   to   $u=+0,\,\, -1<v<1 $: \cite{Clem} 
  \begin{eqnarray}
\varphi (+0,y)&=& -2 \tan^{-1}\frac{v- v_0}{u_0} \nonumber\\
& =&2 \tan^{-1}\frac{\cos \pi y-\cosh \pi x_0\,\cos \pi y_0}{\sinh \pi x_0\,\sin \pi y_0} \,.  
 \label{eq9}
\end{eqnarray}
E.g., for a vortex in the strip middle  $ y_0=1/2 $, we have:
  \begin{eqnarray}
\varphi (+0,y)=2 \tan^{-1}\frac{\cos \pi y }{\sinh \pi x_0 }
  \,.  
 \label{eq10}
\end{eqnarray}
\begin{figure}[htb]
\begin{center}
 \includegraphics[width=8cm] {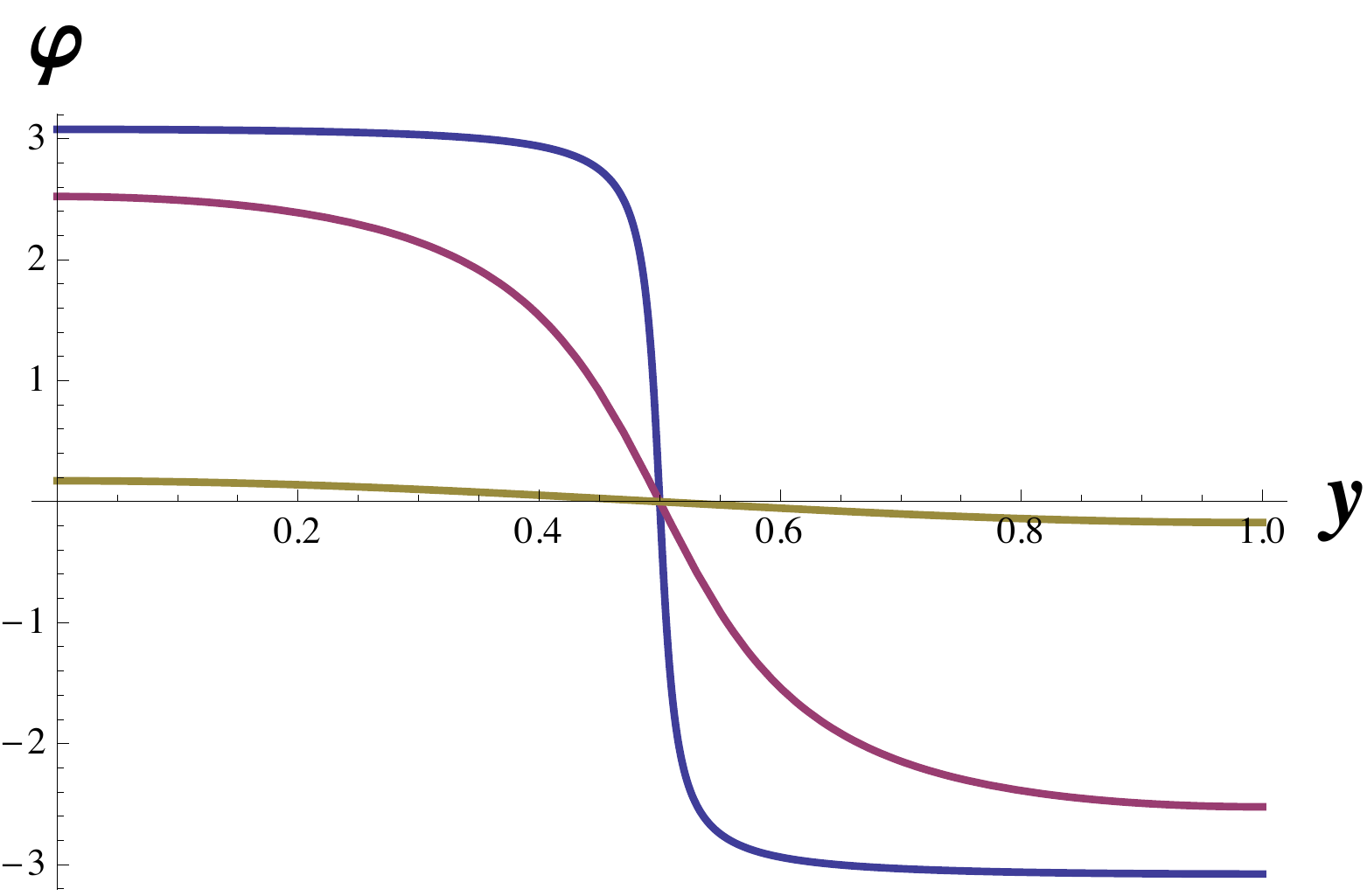}
\caption{(Color online) Contribution of a vortex at $(x_0,1/2)$  to the phase at the right bank of the  junction ($x=+0$) as a function of the   coordinate $y$ along the junction (in units of the junction length $W$). The curves are for $x_0=0.01,\,\, 0.1,\,\, 1$. For $x_0=1$, this contribution is close to zero and nearly constant.   }
\label{fig3}
\end{center}
\end{figure} 
Hence, as is seen from Fig.\,\ref{fig3}, for the vortex close to the junction, $x_0\to 0$, the junction acquires the  $\pi$ character everywhere except a narrow region  of the fast changing vortex phase near $y_0$.\cite{remark} It is seen from the figure that the vortex close to the junction induces the phase difference reminiscent of that of a Josephson vortex. In the latter case, however, the vortex ``core" or the region of the fast phase change is related to a fixed material parameter, the Josephson penetration depth, whereas in our case, as Eqs.\,(\ref{eq9}) and (\ref{eq10}) show, it is of the order $x_0$, the vortex-junction separation.\\

3. $\bm{Far-away\,\, vortex.}$ It is  of interest to examine what happens when the vortex is far from the junction. As  follows from Eq.\,(\ref{eq9}) for large $x_0$, the vortex contribution to the phase at the junction is a constant depending on the vortex position $y_0$:
  \begin{eqnarray}
\varphi (+0 ,y)= \pi ( 2 y_0-1)= C  \,.  
 \label{eq11}
\end{eqnarray}
It is worth noting a remarkable new feature:  the vortex-junction separation $x_0$ drops from this result. 
A straightforward numerical check shows that the last equation approximates the full expression (\ref{eq9})  for $x_0>2$ with accuracy of less than 1\%. Hence, the term ``far-away" can be used reliably for vortices separated from the junction by more than $2W$.
 
When the vortex is in the strip middle, $C=0$, i.e., a far-away vortex does not affect  the junction properties. 
When the vortex is close to the upper edge of the strip, $C=\pi$, whereas  for $y_0\to 0$, the lower strip edge, $C=-\pi$. Hence, when a far-away vortex approaches the strip edges, we in fact have a $\pi$-junction. Moreover, any phase shift from 0 to $\pi$ on the junction can be achieved by changing the vortex position $y_0$ from the middle to the strip edges. This  remarkable topological property  allows one to fine-tune the junction behavior by manipulating positions of far-away vortices. 

This,  at first sight  strange effect can be understood by examining  the  current distribution for a vortex near the edge. When vortex approaches the edge the  current lines are squeezed so that the  component $g_x$ along the edge  increases in a shrinking  space between the vortex core and the edge. Formally, $g_x$ diverges when $y_0\to 0$, see Appendix. Hence,  the phase, the gradient of which is proportional to the current, varies very fast at the part of edge adjacent to the vortex. On the other hand, the total phase change along any contour containing the vortex is $2\pi$. Hence, when the vortex approaches the edge, nearly all available phase change of $2\pi$ happens at the edge section adjacent to the vortex, being close to $-\pi$ on one side of this section and to $+\pi$ on the other. On the rest of the edge, the phase is nearly constant. \\

 4. $\bm{Vortex\,\, in\,\, a\,\, thin\,\, film\,\,  loop. }$ 
Till now we considered an infinite strip with a junction and a vortex in one, e.g.,  the right half-strip. The vortex affects only the phase on the right  junction bank.
If however one has a closed superconducting thin-film loop, the vortex affects the   phases on both junction banks. We do not have an exact solution for the vortex currents in this case. 
Still, one can use a qualitative argument to describe the situation as follows. 


 Consider a straight strip of a finite length $L \gg W$ with no junction and a vortex situated   far from the ends  $x=0, L$. 
 In principle,  this problem can be solved  by conformal mapping, but physically 
 the picture should be similar to the half-infinite strip:  according to Eq.\,(\ref{eq11}) the phase at $x=0$ will be $-\pi (1-2 y_0)$ and at the end $x=L$ it is  $\pi (1-2 y_0)$ (as mentioned, the coordinate $x_0$ of the vortex does not enter these contributions). 

We can now bend this long, but finite, strip to make a ring or loop so that the ends at $x=0$ and $x=L$ come together to form a tunnel junction. The strip bending to a ring should have a small effect if $L\gg W$. Then the junction phase difference  will be close to
  \begin{eqnarray}
\delta\varphi=-\pi (1-2 y_0) -\pi(1-2 y_0)  = -2\pi + 4\pi y_0\,.\qquad  
 \label{eq12}
\end{eqnarray}
$2\pi$ is irrelevant, hence  $\delta\varphi =  4\pi y_0$. This gives $\delta\varphi =  \pi $ for $y_0=1/4$ or $3/4$. Thus, when  the vortex is near the edges $y_0=0,1$ or in the strip middle with $y_0=1/2$, its contribution to the junction phase difference is zero, whereas for the vortex at $y_0=1/4,3/4$ the junction acquires an extra phase difference of $\pi$. Hence, we expect that in a closed superconducting loop with a junction, a vortex at large distances from the junction still affects the junction properties, albeit differently from the case of infinite straight strips.\\
 
5.  $\bm{Discussion.}$
 The effect we describe is due to topologic properties of the  phase which extend to macroscopic distances of the superconducting coherence.  As such it depends on the sample geometry. It should be stressed that in our derivation, the    connection between the phase difference on the Josephson junction and the vortex position is based on the assumption that the thin-film size is small relative to the Pearl length $\Lambda$. This allows one to disregard the magnetic field associated with vortex currents and to deal only with kinetic part of the London energy. Formally, the problem  becomes equivalent to those of  2D electrostatics and the  conformal mapping can be employed. The 3D effects in bulk materials are more complicated since one has to take into account magnetic fields,\cite{Fistul}  the subject out of the scope of this paper. 

Another   property of the mesoscopic system considered here should be mentioned. Since the Josephson energy is $\propto (1-\cos \delta\varphi)$ and the junction phase difference $\delta\varphi$ depends on the vortex position, the total energy of the system differs from the sum of the junction energy with no vortex and the vortex energy with no junction. In other words, there is the junction--vortex interaction  which depends on the vortex position $y_0$ across the strip. Within the model considered, this   interaction is of a purely topologic nature and does not depend on the junction--vortex distance for large separations. \\

 The authors are grateful to J. Clem, A. Gurevich, J. Kirtley, I. Sochnikov, and V. Krasnov  for helpful   discussions. This work was supported by the U.S. Department of Energy, Office of Science, Basic Energy Sciences, Materials Science and Engineering Division. The work was done at the Ames Laboratory, which is operated for the U.S. DOE by Iowa State University under contract  DE-AC02-07CH11358. 

\appendix

\section{Vortex near strip edges}

According to Ref.\,\onlinecite{Pearl}, the magnetic flux crossing the narrow strip due to a vortex at $(0,y_0)$, 
  \begin{eqnarray}
\Phi_z=\frac{ \phi_0}{ \pi \Lambda}\left(y_0\ln \frac{1- y_0}{ y_0} -\ln(1- y_0)\right) \,,  
 \label{a1}
\end{eqnarray}
vanishes as $-y_0\ln y_0$ when the vortex approaches the edge $y=0$ . The question arises  whether or not the sheet current $g_x(y_0\to 0)$ at this edge vanishes as well. 

The current distribution due to the vortex at $(0,y_0)$ is given by the stream function\cite{Pearl}
  \begin{eqnarray}
S=\frac{c\phi_0}{4\pi^2\Lambda}\tanh^{-1}\frac{\sin\pi y\,\, \sin\pi y_0}{\cosh\pi x-\cos \pi y\,\, \cos \pi y_0} \,.  
 \label{a2}
\end{eqnarray}
Consider  $g_x=\partial S/\partial y$ at $x=0$ and $0<y<y_0$:
  \begin{eqnarray}
g_x=\frac{c\phi_0}{4\pi\Lambda} \,\frac{  \sin\pi y_0}{ \cos \pi y -\cos \pi y_0} \,.  
 \label{a3}
\end{eqnarray}
When $y_0\to 0 $ and so does  $y<y_0 $, we obtain:
  \begin{eqnarray}
g_x\propto \frac{1}{y_0-y}\to\infty \,.  
 \label{a4}
\end{eqnarray}
Hence, the phase varies very fast along the part of the edge adjacent to the vortex.

\end{document}